\documentclass[conference]{IEEEtran}
\IEEEoverridecommandlockouts
\usepackage{cite}
\usepackage{amsmath,amssymb,amsfonts}
\usepackage{pxfonts}
\usepackage{algorithmic}
\usepackage{graphicx}
\usepackage{textcomp}
\usepackage{soul}
\def\BibTeX{{\rm B\kern-.05em{\sc i\kern-.025em b}\kern-.08em
    T\kern-.1667em\lower.7ex\hbox{E}\kern-.125emX}}
\begin{document}

\title{Bipartite graph analysis as an alternative to reveal clusterization in complex systems}

\author{
\IEEEauthorblockN{Vasyl Palchykov}
\IEEEauthorblockA{\textit{Laboratory for Statistical Physics of Complex Systems} \\
\textit{Institute for Condensed Matter Physics, NAS of Ukraine}\\
Lviv, Ukraine \\
\textit{$\mathbb{L}^4$ Collaboration \& Doctoral College }\\
\textit{for the Statistical Physics of Complex Systems}, \\
Leipzig-Lorraine-Lviv-Coventry, Europe\\
palchykov@icmp.lviv.ua}
\and
\IEEEauthorblockN{Yurij Holovatch}
\IEEEauthorblockA{\textit{Laboratory for Statistical Physics of Complex Systems} \\
\textit{Institute for Condensed Matter Physics, NAS of Ukraine}\\
Lviv, Ukraine \\
\textit{$\mathbb{L}^4$ Collaboration \& Doctoral College }\\
\textit{for the Statistical Physics of Complex Systems}, \\
Leipzig-Lorraine-Lviv-Coventry, Europe\\
hol@icmp.lviv.ua}
}

\maketitle

\begin{abstract}
We demonstrate how analysis of co-clustering in bipartite networks may be used as a bridge to connect, compare and complement clustering results about community structure in two different spaces: single-mode bipartite network projections.
As a case study we consider 
scientific knowledge, which is represented as a complex bipartite network of articles and related concepts.
Connecting clusters of articles and clusters of concepts via article-to-concept bipartite co-clustering, 
we demonstrate how concept features (e.g. subject classes) may be inferred from the article ones.


\end{abstract}

\begin{IEEEkeywords}
bipartite network, knowledge graph, clustering, modularity, one-mode projections
\end{IEEEkeywords}

\section{Introduction}

Among different types of systems one may distinguish a special class of complex systems \cite{Thurner2017}.
Such systems consist of many interacting parts and these interaction patterns may result in a new level of 
organization in the entire system -- emergent phenomena, which 
appear as a result of bottom-up local interactions 
rather than a centralized top-down control,  see e.g. \cite{Holovatch2016}. 
Consequently, such decentralized system may behave as a new organism;
its behaviour differs significantly from the behaviour of its constituents and is rather governed by connectivity patterns between these parts,
which often exhibit quite complex properties and form a complex network topology \cite{Newman2010,Barabasi2016}.

The systems that belong to a class of complex systems 
may be observed in different environments and include both natural and social phenomena.
Examples are formations in flock of birds or school of fishes, evacuation of crowds, opinion formation in society, financial markets and formation of financial bubbles.
In order to understand, to model or to investigate the scenarios of behaviour for such systems one may need to analyze the underlying network topology.
This include two-step procedure: i) to represent the system as a network or a graph, i.e. a collection of nodes connected by links, and ii) to perform analysis of the corresponding network topology.

The structure of the underlying network may be often expressed as a bipartite graph (or even more complicated), see e.g. \cite{Barabasi2016}. Such networks has two types of nodes. The links connect the nodes of different types and mean the existence of some kind of relations between them. One may think of scientific publications and authors \cite{Newman2001}, public transport routes and serviced stations \cite{Ferber2009}, foods and their ingredients \cite{Ahn2011}, etc. However, these networks are often simplified to one mode projections, e.g. scientific publications connected to each other if they have an author in common, or co-authorship network of researchers. Then these projections are investigated using some tools. For instance community detection or clustering approaches \cite{Fortunato2010} may be used to extract groups or modules in such systems.

One of the goals of this paper is to demonstrate how one-mode projections of the bipartite network may give us different insight into the whole system and how the three representations (two one-mode projections and the bipartite one) may organically supplement one another via clustering analysis.
As a case study 
we will use scientific publication records of manuscripts in a physics domain and extracted concepts \cite{Palchykov2016}.
Besides an obvious demonstration of how bipartite graphs and their projection may be used to extract clustering structure within a system, our article answers a general question concerning organization of scientific knowledge. The rest of the paper is organized as follows: in Section~\ref{sec:Knowledge} we describe two alternative ways to define scientific knowledge. Section~\ref{sec:Data} describes the way the data have been collected, followed by a brief overview of the dataset. In Section~\ref{sec:Networks} we show different ways how to represent data set as a network.
In Section~\ref{sec:Clustering} we present clustering analysis of these networks and summarize the results in Section~\ref{sec:Conclusions}.


\section{Knowledge: alternative approaches}\label{sec:Knowledge}
There is no strict and unique mathematical formalization of scientific knowledge. 
Let us assume that each original scientific publication produces a new piece of knowledge. These pieces of knowledge do not ``live'' in isolation, but are connected to the other pieces. They may be connected to existing pieces of knowledge by citation links. Thus, one arrives at a simplified view of the knowledge as a citation graph \cite{Price1965}. The nodes of such graphs correspond to scientific publications and chronologically directed links indicate citation links between papers. Besides citation graph one may consider other types of relationships between scientific publications, e.g. content or author-based similarity metrics, see e.g. \cite{Waltman2010}.

Alternative view of the knowledge (and its constituents) assumes that scientific publications consist of the pieces of knowledge rather than represent such pieces themselves. Here scientific ideas may be considered as constituents of knowledge, which may be represented by concepts in scientific publications. Each publication may contain a number of concepts. Examples of scientific concepts in physics domain of science include \texttt{Center of mass}, \texttt{Momenta of inertia}, \texttt{Conservation of energy}, etc. Once a new concept appears in a scientific communication, it is assumed that the (scope of) existing knowledge has been extended, see e.g. \cite{Iacopini2018}. 

To use this approach, scientific concepts have to be extracted from the bodies of scientific publications. In the next chapter we describe the process of such concept extraction.

\section{Data and Exploratory Analysis}\label{sec:Data}

It is possible to obtain basic or generic set of concepts in scientific domain, but it is much more complicated to get a comprehensive vocabulary of concepts even in a single domain of science due to the following reasons: i) constant evolution of concepts due to the development of scientific knowledge, and ii) fragmentation of science. It is natural to assume that a comprehensive vocabulary of scientific concepts may be possessed only by active  scientist, and only in the field of his or her research. In order to create up-to-date vocabulary (ontology) of scientific concepts it may be required to combine {\em computational capabilities} with the efforts of researchers from different domains and regularly (if needed) upgrade it.
With this purpose in mind \texttt{ScienceWISE} (SW) platform has been deployed \cite{Astafiev2012,Martini2016}. This platform was designed to allow scientists to navigate over scientific literature, using scientific concepts as ``labels'' (or ``tags'') for navigation. Initial vocabulary of concepts has been obtained using a semi-automated import from science-oriented ontologies and online encyclopedia \cite{Astafiev2012}. Afterwards the users of SW were allowed to edit ontology to make navigation more reliable. For instance if a user observes that some concept (e.g. \texttt{bipartite network}) is missed, and the concept will help to navigate over the literature, he or she may add the concept to the vocabulary. Once a new concept is added to the vocabulary, all the articles have to be re-scanned in order to upgrade concept list for each paper.

Below we investigate the structure of concept-related networks of scientific knowledge. A primary source of literature for SW platform is \texttt{arXiv} e-print repository of manuscript \cite{Ginsparg2011},  which allows for the full-text access to all manuscripts even before they are officially published.

Here we use scientific concepts extracted from research publications (\texttt{arXiv} preprints) using the SW platform. These concepts have been previously investigated in \cite{Palchykov2016, Martini2016, Martini2017}. Here we are interested in investigation of the static picture of concept (knowledge) network, thus we restrict ourselves to a single year, namely 2013. Another reason is that some structural properties of the corresponding network of publications (on the same data sets) have been investigated in \cite{Palchykov2016}. 
Here we again restrict our analysis to the manuscripts to which a single category has been attached by the authors. 
It is worth to mention that during manuscript submission process to \texttt{arXiv} the authors are required to classify their manuscript, i.e. assign it to at least one subject class in \texttt{arXiv} classification scheme. This classification scheme consists of thirteen subject classes, which include in particular astrophysics (\texttt{astro-ph}), condensed matter physics (\texttt{cond-mat}), four classes of high energy physics (\texttt{hep-ph}, \texttt{hep-ex}, \texttt{hep-lat} and \texttt{hep-th}), two classes of nuclear physics (\texttt{nucl-th} and \texttt{nucl-ex}), general relativity and quantum cosmology (\texttt{gr-qc}), physics (\texttt{physics}), quantum physics (\texttt{quant-ph}), nonlinear sciences (\texttt{nlin}) and mathematical physics (\texttt{math-ph}).

The subject of our analysis was a collection of manuscripts submitted during the year 2013 that accounts for 36386 articles. These articles contain 12200 unique concepts in common. 347 of these concepts have an expert given \texttt{generic} label, which means that the concept has a generic meaning in physics. Examples include \texttt{Energy} or \texttt{Field}. Each manuscript contains on average $37$ unique non-generic concepts with significant differences among the articles: the number of identified concept within an article varies between 1 and $\approx400$. A brief summary of the data set properties is shown in Table~\ref{tab_00}.
\begin{table}[htbp]
\caption{Basic characteristics of the dataset: total number of manuscripts ($N$), total number of identified concepts ($V$) and the number of generic ones ($V_{\rm gen}$) among them; $\langle{k}\rangle$ stands for the average number of non-generic concepts in an article.}
\begin{center}
    \begin{tabular}{|c|c|c|c|c|}
 \hline
			&$N$		&$V$	&$V_{\rm gen}$	&$\langle{k}\rangle$\\\hline
\texttt{arxivPhys2013}	&$36386$&	$12200$&	$347$&	$37$\\\hline
\end{tabular}
\label{tab_00}
\end{center}
\end{table}

\section{Network representations}\label{sec:Networks}

\begin{figure}[!ht]
\begin{center}
\includegraphics[width=8cm]{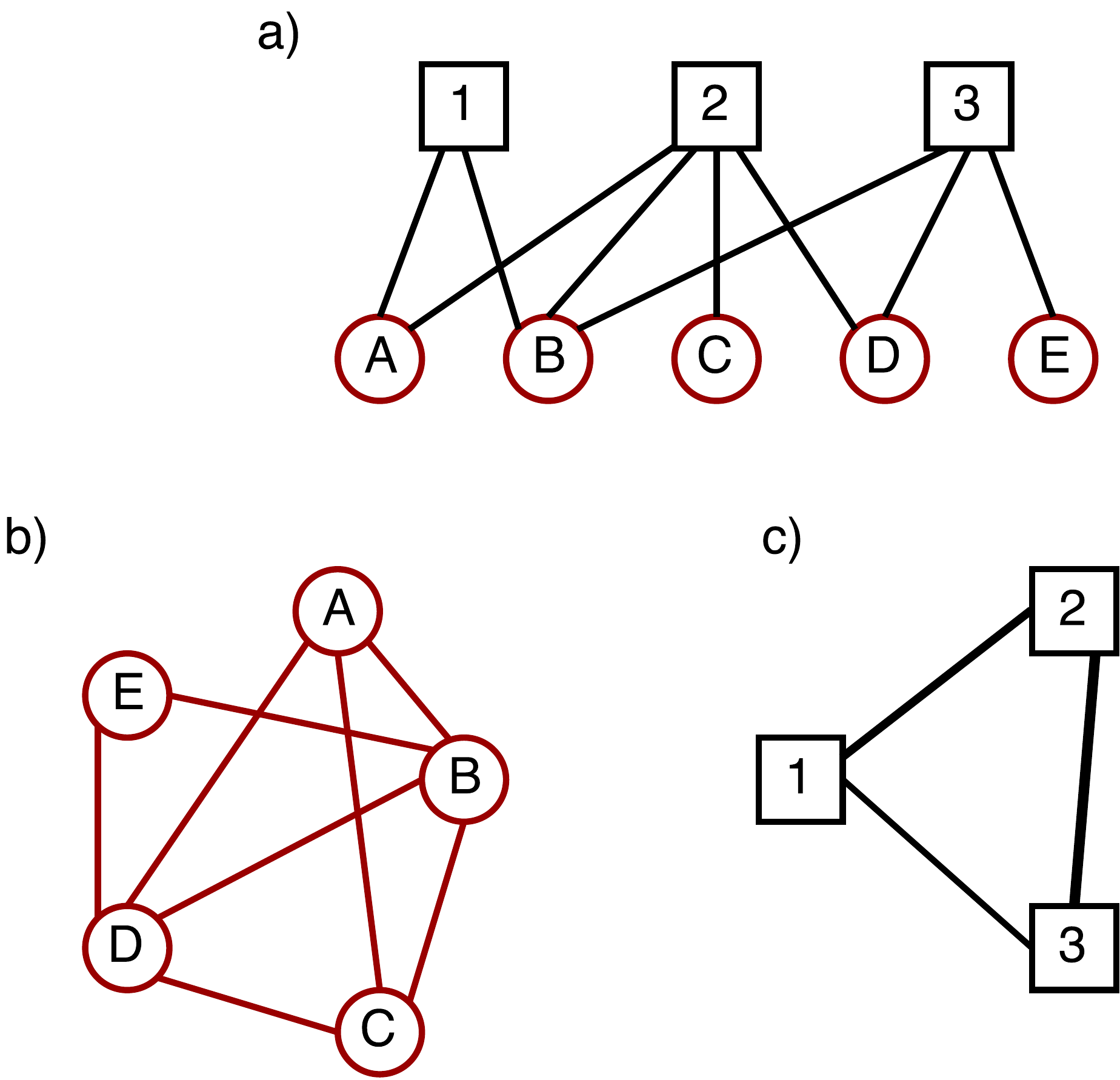}
\caption{Examples of network representations for scientific publication system. Panel {\bf a)}: unweighted bipartite representation that contains two types of nodes: articles (squares) and concepts (circles) and the links connecting the nodes of distinct types. Two other panels represents one-mode projections of the bipartite network. Panel {\bf b)}: concept network that consists only of concept-nodes; two nodes are connected in the corresponding concepts co-occurred together at least in one publication. 
Panel {\bf c)}: the nodes of the network represent scientific articles and two article-nodes are connected if they used the same terminology (they have at least one concept in common). This network may be also called \texttt{content (concept) coupling network} in analogy to the \texttt{bibliographic coupling network}. 
\label{fig:nets}
}
\end{center}
\end{figure}

As a basic network representation of the publication system we consider bipartite network. Here manuscripts and the identified concepts are mapped into two types of network nodes.
A link connects an article-node and a concept-node if the corresponding concept has been identified within the text of the article. The corresponding representation is shown in Fig.~\ref{fig:nets}\textbf{a}.

There are two possible one-mode projections of this bipartite network. Considering only concept-nodes and linking each pair of these nodes that co-occurred in (at least once) the same publication, we arrive at the projection to the \texttt{concept space}, see Fig.~\ref{fig:nets}\textbf{b}. For simplicity all links are assumed to have the same (unit) weight. 

Alternatively, one may build a supplementing projection to the \texttt{article space}. Here two article-nodes are connected to each other if they use at least one concept in common. Since such networks are extremely dense (there are over 50\% of all possible links in a network), a weighted networks configuration is considered here. Firstly we represent each article $i$ as a vector $a_i$ in a multi-dimensional space, where each dimension corresponds to each of $V-V_{\rm gen}$ unique concept.
 A weight $w_{ij}$ of a link between articles $i$ and $j$ is defined as a cosine between the corresponding article vectors:
 \begin{equation}
 w_{ij} = \cos(\vec{a_i}, \vec{a_j}).
 \end{equation}
 To account for the heterogeneity in the widespread of concepts, each concept is weighted according to its \texttt{idf} factor:
 \begin{equation}
 \texttt{idf}(c) = \log \frac{N}{N_c},
 \end{equation}
 where $N_c$ is the number of articles in collection that contain concept $c$. Thus, defining $\vec{a_i}$ such that
 \begin{equation}
  a_{i,c} =
  \begin{cases}
    \texttt{idf}(c), & \text{if $c \in i$} \\
    0, & \text{otherwise}
  \end{cases}
\end{equation}
we take into account that the usage of a common widespread concept affects the similarity score between two articles less than the usage of a more specific one.

%

\section{Modular structure}\label{sec:Clustering}

Once the networks are generated, let us investigate their community structure.
In order to identify clusters in a network, modularity \cite{Newman2004} optimization approach has been applied, with the Barber's modification \cite{Barber2007} for bipartite networks.
To maximize modularity, greedy optimization Louvain algorithm \cite{Blondel2008} has been applied. Due to its stochastic nature, which results in a local rather than global maximum at each run, we performed 100 runs on unipartite networks and 1000 runs on a bipartite network. 
Then a single partition for each network with the highest value of modularity has been chosen.
Note that the clusters of one-mode projections consist of the nodes of a single type only, while the cluster of a bipartite network, in general, consists of both article nodes and concept-nodes.
The scores of the highest modularity $M$ ($M_{\rm B}$ for bipartite network) obtained:
\begin{itemize}
  \item \texttt{Bipartite network}: $M_{\rm B}=0.453262$
  \item \texttt{Concept network}: $M=0.245219$
  \item \texttt{Article network}: $M=0.329519$
\end{itemize}
In what follows below we consider only the clusters that have more than one node (unipartite networks) or more than two nodes (bipartite network).

Previously \cite{Palchykov2016} the clusters of scientific manuscripts have been compared to their expert made classification, both for bipartite network and its projection to article space. Since each cluster of a bipartite partition consists of a set of articles and concepts, a subset from each cluster, namely the articles that fall into this cluster, have been considered.
The results demonstrated: i) similarity between the obtained clusters and author made classification, and ii) some discrepancies between the two. The detailed analysis showed us that some of these discrepancies have underlying reasons behind, indicating both historical classification reasons and methodological similarity between rather unrelated domains.
It appeared \cite{Palchykov2016} that the bipartite network contained six clusters and the article-to-article one contained four clusters, which provided rather similar results.

\begin{figure}[!ht]
\begin{center}
\includegraphics[width=8cm]{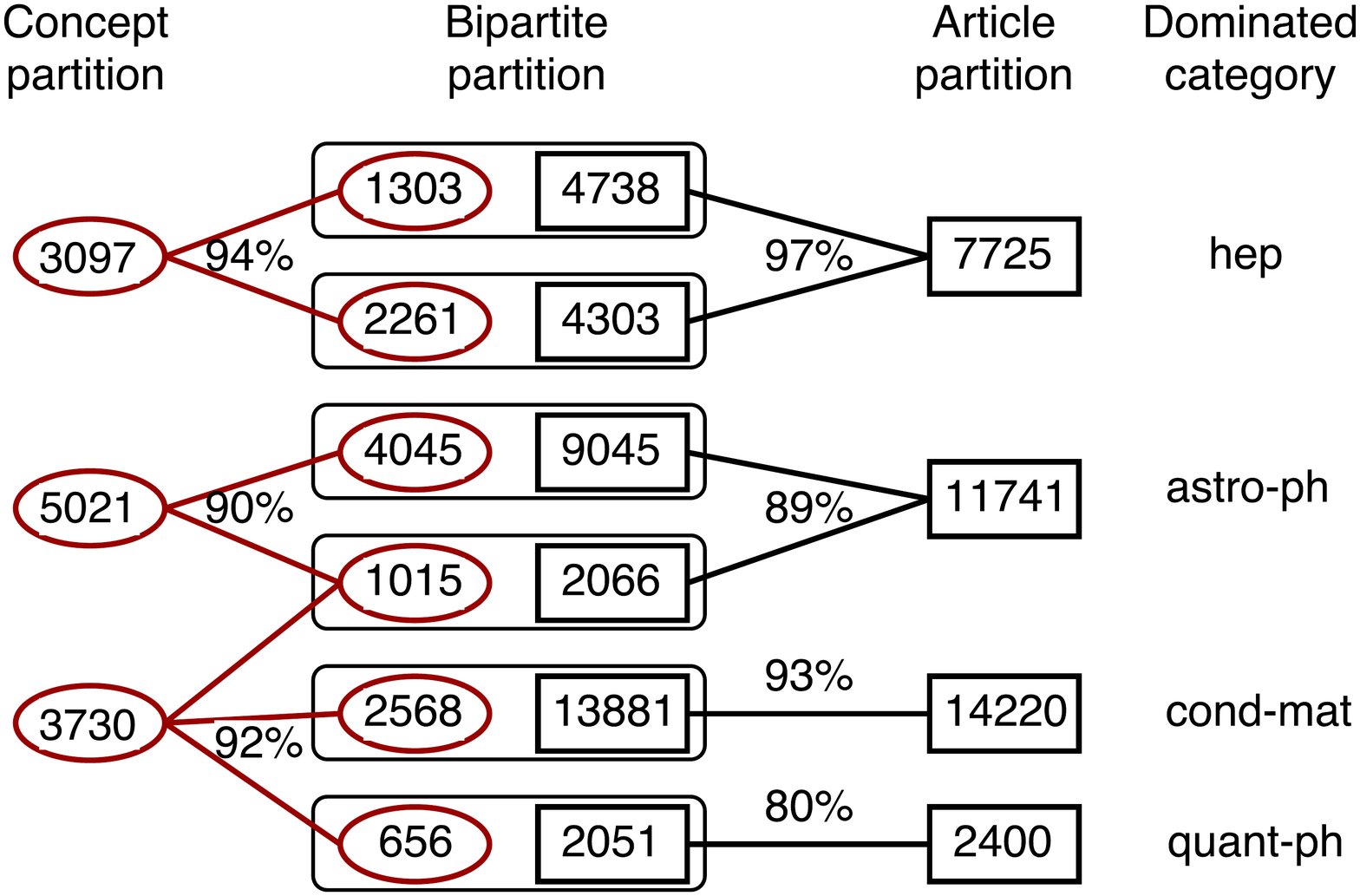}
\caption{Clusters of concept network (left hand side, ovals), bipartite network (center) and a article network (right hand side, rectangles). The number in an ovals gives the number of concepts in the cluster, the number in a rectangle gives the number of articles in the cluster.
A pair of numbers in a bipartite network gives the number of concepts and the number of articles in each cluster of a bipartite network. A category code next to each article cluster denotes the category to which most of the articles of the cluster belong. Percentage near each cluster of an article or a concept network show the fraction of nodes in the cluster that belongs to a linked cluster(s) of the bipartite network.}
\label{fig:clusters}
\end{center}
\end{figure}

In the present research we extend the analysis of Ref. \cite{Palchykov2016} by adding the concept dimension: instead of ignoring the concepts that fall within each bipartite cluster we take them into account and use the bipartite combined clusters as a bridge between pure partitions of article and concept unipartite projection partitions.
To illustrate this let us mention that unlike articles, scientific concepts lack expert made classification.
So, we cannot directly assign arXiv subject classes to the clusters of concepts. This, however, may be done using combined (consisting of concepts and articles) partitions of bipaprtite clusters, as illustrated in Fig.~\ref{fig:clusters}.

Fig.~\ref{fig:clusters} displays a brief description of the optimal partitions in bipartite network and its both projections together with some relations. 
Beside six clusters of the bipartite network optimal partition, and four ones of the article-to-article network, the optimal partition of the concept network consists of three clusters.
Let us first consider the first two clusters of bipartite networks (shown at the top of Fig.~\ref{fig:clusters}). The articles that belong to these clusters are dominated by high energy physics (\texttt{hep-}) categories.
The main difference between these two clusters is that the first one is rather focussed on experimental observations, while the second cluster is more about theoretical approaches to the problem, for details see \cite{Palchykov2016}. 
In the unipartite projection to the article space these two clusters rather form a single one (at the top), which we may call \texttt{article:hep}. 
Moreover there is quite good correspondence between these clusters: 97\% of articles of \texttt{article:hep} cluster belong to either of the two clusters of the bipartite partition. 
On the other hand, there is a good correspondence between these clusters and the top cluster (Fig.~\ref{fig:clusters}) of a unipartite projection to concept space: 94\% of all concepts that fall into this cluster belong to the considered clusters of bipartite partition. Thus, we call this cluster as \texttt{concept:hep}. These are the concepts that has dominant usage in high energy physics research.

Similar picture may be observed for \texttt{astro-ph} subject class. 89\% of nodes of the second cluster in article space belong either to the third or to the fourth cluster in bipartite space, and again the concepts of the latter clusters form 90\% of all concepts belonging to the second cluster in concept space. Thus, as a good approximation, the concepts belonging to this cluster may be labeled as \texttt{concept:astro-ph} ones.

The situation with \texttt{cond-mat} and \texttt{quant-ph} subject classes is different. While they form rather separated clusters in article space, the concepts used in these subject classes fall into a single cluster. These observations hint that besides considering different objects and on different scales the methodology behind these subject classes may overlap significantly.

\section{Conclusions}\label{sec:Conclusions}
The structure of many complex systems can be expressed in terms of the underlying bipartite network, which connects different types of components. Investigation of such systems is often done considering one mode projections of these separately. 
Examples include co-authorship network of paper-to-author bipartite graph \cite{Newman2003}, etc. To find patterns in such networks one may apply a variety of tools including clustering algorithm to identify groups of tightly related items in a system. Here we show how co-clustering of bipartite network may be used as a bridge to connect and complement clustering results in two different projections. Considering scientific publications in physics domain and a set of extracted concepts from their texts, we build bipartite article-to-concept networks and made its both one mode projections. We show how the information about one part of the system may add value to the other one. In the considered case publications possess the author made classification according to arXiv subject classes, however scientific concepts lack such classification. By comparing such concepts we were able to assign such classes to the concepts. Moreover such approach allows us to compare groups of completely different objects: articles and concepts. We see a hint that two different subject classes, \texttt{cont-mat} and \texttt{quant-ph}, even though being well distinguishable in terms of scientific publications, use terminology (concepts), which is quite similar.

\section*{Acknowledgment}
This work was supported in part by FP7 EU IRSES project No. 612707 ``Dynamics of and in Complex Systems''
and by the Ukrainian DFFD by the project 76/105-2017 (Yu.H.).

\end{document}